\documentclass[12pt]{iopart}

\usepackage{graphicx}

\usepackage{iopams}

\newtheorem{theorem}{Theorem}

\newtheorem{proposition}{Proposition}

\newenvironment{proof}[1][Proof]{\textbf{#1.} }{\ \rule{0.5em}{0.5em}}

\begin{document}

\title[]{Determinantal formulae for the Casimir operators of inhomogeneous Lie algebras}

\author{Rutwig Campoamor-Stursberg\dag}

\address{\dag\ Dpto. Geometr\'{\i}a y Topolog\'{\i}a\\Fac. CC. Matem\'aticas\\
Universidad Complutense de Madrid\\Plaza de Ciencias, 3\\
E-28040 Madrid, Spain}

\ead{rutwig@mat.ucm.es}

\begin{abstract}
Contractions of Lie algebras are combined with the classical
matrix method of Gel'fand to obtain matrix formulae for the
Casimir operators of inhomogeneous Lie algebras. The method is
presented for the inhomogeneous pseudo-unitary Lie algebras
$I\frak{u}(p,q)$. This procedure is extended to contractions of
$I\frak{u}(p,q)$ isomorphic to an extension by a derivation of the
inhomogeneous special pseudo-unitary Lie algebras
$I\frak{su}(p-1,q)$, providing an additional analytical method to
obtain their invariants. Further, matrix formulae for the
invariants of other inhomogeneous Lie algebras are presented.
\end{abstract}



\maketitle

\section{Introduction}

Unitary Lie algebras constitute one of the most important classes
of Lie algebras appearing in physics, as shows their application
in hadron physics, the quark model or spontaneously broken
symmetries. In combination with other Lie algebras, pseudo-unitary
algebras are also relevant for physical phenomena, like the
quaplectic group in the Born reciprocity. Generalized
In\"on\"u-Wigner contractions of these algebras are therefore also
of interest, because they allow to relate certain model
transitions, as happens for the quantum mechanical versions of the
liquid drop model \cite{Gi}, where models exist whose dynamical
group is respectively the unitary group $U(6)$ and the
inhomogeneous unitary group $IU(5)$. Since the latter can be
obtained by means of In\"on\"u-Wigner contractions of $U(6)$, the
corresponding models are related by a limiting process.
Inhomogeneous unitary algebras have also been used to describe
quantum states in relativistic theories of internal particle
properties \cite{Pj}. In all these problems, effective expressions
for the Casimir operators of the unitary and inhomogeneous unitary
algebras are needed. In contrast to the simple or reductive
algebras, the computation of the invariants of inhomogeneous
algebras $\frak{g}$ is not feasible using the classical theory,
and various approaches have been developed \cite{De2,Hp,Pe,Ca3},
using either the universal enveloping algebra $\frak{U}(\frak{g})$
explicitly or by means of analytical reductions of differential
equations.

In this work we analyze the  generalization of the matrix
procedure of Gel'fand \cite{Ge} to various inhomogeneous Lie
algebras. This generalization does not involve computations in the
enveloping algebras or reductions of differential equations, but
focuses on the combination of the Gel'fand method for simple Lie
algebras with contractions of Lie algebras \cite{He,Lo}. We first
compute the Casimir operators of inhomogeneous pseudo-unitary
algebras $I\frak{u}(p,q)$ by means of characteristic polynomials,
and then extend the result to certain of its contractions. We also
analyze the corresponding problem for inhomogeneous special
pseudo-unitary algebras $I\frak{su}(p,q)$, which are obtained
contracting\footnote{More specifically, as a direct summand of a
contraction.} either $\frak{su}(p+1,q)$ or $\frak{su}(p,q+1)$
\cite{Ro}. However, here the contraction of the corresponding
Casimir operators leads to dependence problems. To avoid this
situation, we construct a special contraction $K\frak{su}(p,q)$ of
$\frak{su}(p+1,q)$, which can be realized as an extension by a
derivation of the inhomogeneous algebras $I\frak{su}(p,q)$. We
give a determinantal formula for the invariants of this
contraction, and determine a maximal set $\left\{W_{k}\right\}$ of
invariants which satisfy a certain constraint. Analyzing how these
functions $W_{k}$ transform when inserted in the system of PDEs
giving the invariants of $I\frak{su}(p,q)$, we propose a simple
analytical method to compute Casimir operators of the latter
algebra. This will moreover explain why contraction of
$\frak{su}(p,q)$-invariants leads to powers of the quadratic
Casimir operator of $I\frak{su}(p,q)$, as already observed in
\cite{Ro}. We finally present matrix formulae to determine the
Casimir invariants of other Lie algebras, namely the inhomogeneous
Lie algebras $I\frak{so}(p,q)$ and two of its contractions that
can be seen as a generalization of the classical Galilei and
Carroll Lie algebras.

\bigskip

The method to obtain the invariants of a Lie algebra that we will
use in this work is the analytical approach, i.e., using  the
representation of $\frak{g}$ by differential operators
\begin{eqnarray}
\widehat{X}_{i}=-C_{ij}^{k}x_{k}\frac{\partial}{\partial{x_{j}}},\quad 1\leq i\leq n,
\end{eqnarray}
where $\left\{C_{ij}^{k}\right\}$ is the structure tensor over the
basis $\left\{X_{1},..,X_{n}\right\}$ and
$\left\{x_{1},..,x_{n}\right\}$ the dual basis. A function
$F(x_{1},..,x_{n})$ is an invariant if it satisfies the system:
\begin{equation}
\widehat{X}_{i}F(x_{1},..,x_{n})=-C_{ij}^{k}x_{k}\frac{\partial }{\partial x_{j}}F\left(
x_{1},..,x_{n}\right) =0, \quad 1\leq i\leq n \label{sys}
\end{equation}
If $F$ is a polynomial solution of (\ref{sys}), its symmetrization
corresponds to a classical Casimir operator (see e.g. \cite{De2}).
The number $\mathcal{N}(\frak{g})$ of independent solutions of the
system (non necessarily polynomials) is given by \cite{Be}:
\begin{equation}
\mathcal{N}(\frak{g})=\dim \,\frak{g}- {\rm rank}\left( C_{ij}^{k}x_{k}\right),
\end{equation}
where $A(\frak{g}):=\left(C_{ij}^{k}x_{k}\right)$ is  the matrix
which represents the commutator table over the basis
$\left\{X_{1},..,X_{n}\right\}$.

We also recall the elementary notions about contractions that will
be needed here. Let $\frak{g}$ be  a Lie algebra and $\Phi_{t}\in
Aut(\frak{g})$ a family of automorphisms of $\frak{g}$, where
$t\in\mathbb{N}$. For any $X,Y\in\frak{g}$ define
\begin{equation}
\left[X,Y\right]_{\Phi_{t}}:=\left[\Phi_{t}(X),\Phi_{t}(Y)\right]=\Phi_{t}(\left[X,Y\right]).
\end{equation}
Obviously $\left[X,Y\right]_{\Phi_{t}}$ are the brackets of the
Lie algebra over the transformed basis. Now suppose that the limit
\begin{equation}
\left[X,Y\right]_{\infty}:=\lim_{t\rightarrow \infty}\Phi_{t}^{-1}\left[\Phi_{t}(X),\Phi_{t}(Y)\right] \label{Ko}
\end{equation}
exists for any $X,Y\in\frak{g}$. Equation (\ref{Ko}) defines a Lie
algebra $\frak{g}^{\prime}$ called the contraction of $\frak{g}$
(by $\Phi_{t}$). If $\frak{g}$ and $\frak{g}^{\prime}$ are
nonisomorphic, then the contraction is called non-trivial. This
procedure can be enlarged to contract Casimir invariants. If
$F(X_{1},..,X_{n})$ is a Casimir operator of degree $p$ of
$\frak{g}$, we can rewrite it in the new basis
$\left\{\Phi_{t}(X_{1}),..,\Phi_{t}(X_{n})\right\}$ and consider
the limit
\begin{equation}
F^{\prime}(X_{1},..,X_{n}):=\lim_{t\rightarrow\infty}t^{p}F(\Phi_{t}(X_{1}),..,\Phi_{t}(X_{n})).\label{Kof}
\end{equation}
It is straightforward to verify, using (\ref{Ko}), that
$F^{\prime}(X_{1},..,X_{n})$ is a Casimir operator of the
contraction. This method has become standard in the literature to
obtain Casimir invariants of contractions \cite{He,Lo,Ca24,Ca33}.

\section{Inhomogeneous pseudo-unitary Lie algebras and their invariants.}

In this work we will use the basis of $\frak{u}(p,q)$ given by the
operators $\left\{ E_{\mu\nu},F_{\mu\nu}\right\}
_{1\leq\mu,\nu\leq p+q=n}$ with the constraints
$E_{\mu\nu}+E_{\nu\mu}=0,\;F_{\mu\nu}-F_{\nu\mu}=0$.  The
commutation relations over this basis are:
\begin{eqnarray}
\left[  E_{\mu\nu},E_{\lambda\sigma}\right]   &  =g_{\mu\lambda}E_{\nu\sigma
}+g_{\mu\sigma}E_{\lambda\nu}-g_{\nu\lambda}E_{\mu\sigma}-g_{\nu\sigma
}E_{\lambda\mu}\label{b1}\\
\left[  E_{\mu\nu},F_{\lambda\sigma}\right]   &  =g_{\mu\lambda}F_{\nu\sigma
}+g_{\mu\sigma}F_{\lambda\nu}-g_{\nu\lambda}F_{\mu\sigma}-g_{\nu\sigma
}F_{\lambda\mu}\\
\left[  F_{\mu\nu},F_{\lambda\sigma}\right]   &  =g_{\mu\lambda}E_{\nu\sigma
}+g_{\nu\lambda}E_{\mu\sigma}-g_{\nu\sigma}E_{\lambda\mu}-g_{\mu\sigma
}E_{\lambda\nu}\label{b2}%
\end{eqnarray}
where $\left(g_{\mu\nu}\right)=\left(1,..,1,-1,..,-1\right)$ is a
 diagonal matrix. Since $\frak{u}\left(  p,q\right)  =\frak{su}\left(
p,q\right) \oplus\mathbb{R}$, it follows at once that
$\frak{u}\left( p,q\right)  $ has $\left(  p+q\right)  $
independent Casimir operators, one of them being
$g^{\mu\mu}F_{\mu\mu}$, for corresponding to the central element,
while the other $(p+q-1)$ invariants correspond to the simple part
\cite{Ra,Po1,Ca42}.

\begin{proposition}
For $N=p+q\geq2$, a maximal set of independent Casimir invariants
of $\frak{u}\left(  p,q\right)  $ is given by the coefficients
$C_{k}$ of the characteristic polynomial $\left|iA_{p,q}-\lambda
{\rm Id}_{N}\right|
=\lambda^{N}+\sum_{k=1}^{N}C_{k}\lambda^{N-k}$, where
\begin{equation}
A_{p,q}=\left(
\begin{array}
[c]{ccccc}%
-if_{11} & .. & -g_{jj}(e_{1j}+if_{1j}) & .. & -g_{NN}\left(  e_{1N}%
+if_{1N}\right)  \\
: &  & : &  & :\\
e_{1j}-if_{1j} & .. & -ig_{jj}f_{jj} & .. & -g_{NN}\left(  e_{jN}%
+if_{jN}\right)  \\
: &  & : &  & :\\
e_{1N}-if_{1N} & .. & g_{jj}\left(  e_{jN}-if_{jN}\right)   & .. &
-ig_{NN}f_{NN}%
\end{array}
\right)  \label{ST}
\end{equation}
and $i=\sqrt{-1}$. Moreover, $\deg(C_{k})=k$ for $k=1..p+q=N$.
\end{proposition}

In particular, if $g_{ii}=1$ for all $i$, we obtain the unitary
algebra $\frak{u}\left(  n\right)$. Using the notation
$a^{ij}=\left( e_{ij}+if_{ij}\right)  $ with the constraint
$a^{ij}=-a^{ji\ast}$, it was shown in [18, chapter 7, p.239] that
the Casimir operators of $\frak{u}\left(  n\right)  $ are obtained
from
the secular equation%
\begin{equation}
\det\left\|  a^{ij}J_{ij}-\lambda I_{n}\right\|  =\left(
-\lambda\right) ^{n-r}\phi_{r}\left(  a^{ij}\right)  ,
\end{equation}
where
\begin{equation}
\phi_{r}\left(  a^{ij}\right)  =\frac{1}{r!\left(  n-r\right)  !}%
\varepsilon_{i_{1}..i_{n}}a^{i_{1}j_{1}}...a^{i_{r}j_{r}}\varepsilon
_{j_{1}..,j_{r}}.
\end{equation}
Here $J_{ij}$ is the fundamental matrix having entry 1 at position
$\left( i,j\right)  $ and zero elsewhere, and $\varepsilon$ the
skew Levi-Civita tensor. Formula (\ref{ST}) is nothing but the
corresponding adaptation to the different real forms of
$\frak{u}(N)$ over the basis (\ref{b1})-(\ref{b2}).

There is an interesting consequence of this formulation, namely,
that the determinant of $A_{p,q}$ satisfies the following identity
\begin{equation}
\det(A_{p,q})=\prod_{j=2}^{N}g_{jj} \det(A_{N,0}).
\end{equation}
This implies that all pseudo-unitary algebras $\frak{u}(p,q)$ have
the Casimir invariant of order $N$ in common, while those of lower
degree depend essentially on the index and signature of the matrix
$(g_{\mu\mu})$.

\bigskip

As shown first by Rosen, inhomogeneous unitary Lie algebras
$Iu\left(  p,q\right)  $ can be obtained from an
In\"{o}n\"{u}-Wigner contraction of the unitary algebras
$\frak{u}\left(  p+1,q\right)  $ or $\frak{u}\left(  p,q+1\right)
$. We will use also this contraction to obtain the matrix formulae
giving the Casimir operators of $I\frak{u}\left( p,q\right)  $,
but without working explicitly with elements of the enveloping
algebra. Our approach uses only matrices, and the invariants will
follow from characteristic polynomials or combinations of them. To
this extent, we must first fix some notation. We consider the
ordered basis $\left\{  F_{ii},E_{ij},F_{ij}\right\}  $ for 1$\leq
i,j\leq N$. Without loss of generality, we can suppose that for
the matrix (\ref{ST}) we have $g_{11}=1$.
We consider the automorphism $\Phi$ of $\frak{u}\left(  p,q\right)  $ given by%
\begin{equation}
\fl
\begin{array}
[c]{lll}%
\Phi\left(  F_{11}\right)  =\frac{1}{t}F_{11}; & \Phi\left(
F_{\mu\mu }\right)  =F_{\mu\mu} \;(\mu\geq2); & \Phi\left(
E_{1\sigma}\right)  =\frac{1}{t}E_{1\sigma}\; (\sigma\geq 2) \\
\Phi\left(F_{1\sigma}\right)  =\frac{1}{t}F_{1\sigma}
\;(\sigma\geq2) & \Phi\left(  E_{\mu\nu}\right)  =E_{\mu\nu}, &
\Phi\left( F_{\mu\nu}\right) =F_{\mu\nu},\;\mu,\nu\neq 1.
\end{array} \label{K1}
\end{equation}
For $\mu,\nu\neq1$ the generators $\Phi\left(  E_{\mu\nu}\right) $
and $\Phi\left(  F_{\mu\nu}\right)  $ obviously generate the
unitary Lie algebra $\frak{u}\left(  p-1,q\right)  $. The limit
$\lim_{t\rightarrow\infty}\Phi^{-1}\left[ \Phi(X),\Phi(Y)\right] $
exists for any pair of elements $X,Y\in\frak{u}\left( p,q\right)
$, and clearly defines a contraction. If we write $\Phi\left(
F_{1\rho}\right) :=Q_{\rho}$ and $\Phi\left( E_{1\rho }\right)
:=R_{\rho}$, then we have the brackets
\begin{equation}
\begin{tabular}{ll}
$\left[ E_{\mu \nu },R_{\rho }\right] =g_{\mu \rho }R_{\nu
}-g_{\nu \rho }R_{\mu },$ & $\left[ E_{\mu \nu },Q_{\rho }\right]
=g_{\mu \rho }Q_{\nu
}-g_{\nu \rho }Q_{\mu },$ \\
$\left[ F_{\mu \nu },Q_{\rho }\right] =-g_{\nu \rho }R_{\mu
}-g_{\mu \rho }R_{\nu }$, & $\left[ F_{\mu \nu },R_{\rho }\right]
=g_{\mu \rho }Q_{\nu
}+g_{\nu \rho }Q_{\mu },$%
\end{tabular} \label{Dar}
\end{equation}
and this coincides exactly with the standard representation of $\frak{u}%
\left(  p-1,q\right)  $ as given in \cite{Ro}. Observe that in particular
$\Phi\left(  F_{11}\right)  $ commutes with any other generator. Therefore the
contraction is isomorphic to the direct sum $I\frak{u}\left(  p-1,q\right)  \oplus\mathbb{R}%
$.

\medskip

Let $N+1=p+q+1\geq 2$ and consider the contraction $\frak{u}(p,q)
\rightsquigarrow I\frak{u}(p-1,q)\oplus <F_{11}>$ determined by
(\ref{K1}). We next give with an analogous formula for the
inhomogeneous Lie algebra $I\frak{u}(p-1,q)$.

\begin{proposition}
In the preceding conditions, a maximal set of independent Casimir
operators $D_{k}$ of $I\frak{u}\left( p-1,q\right) $ is obtained
from the determinant
$\Delta(\lambda)=\sum_{k=2}^{N}D_{k}\lambda^{N-k}$ defined by
\begin{equation}
\fl \lim_{t\rightarrow \infty }\frac{1}{t^{2}}\left|
\begin{array}{ccccc}
f_{11}t-\lambda  & ... & -g_{jj}\left( ie_{1j}-f_{1j}\right) t &
... &
-g_{N+1,N+1}\left( ie_{1,N+1}-f_{1,N+1}\right) t \\
: &  & : &  & : \\
\left( ie_{1j}+f_{1j}\right) t & ... & g_{jj}f_{jj}-\lambda  & ...
&
-g_{N+1,N+1}\left( ie_{j,N+1}-f_{j,N+1}\right)  \\
: &  & : &  & : \\
(ie_{1,N+1}+f_{1,N+1})t & ... & g_{jj}\left(
ie_{j,N+1}+f_{j,N+1}\right)  & ... &
g_{N+1,N+1}f_{N+1,N+1}-\lambda
\end{array}
\right| . \label{Def}
\end{equation}
\end{proposition}

The preceding formula is obtained starting from (\ref{ST}) and
implementing the contraction to the algebra
$I\frak{u}(p,q)\oplus\mathbb{R}$, scaling the determinant to avoid
divergencies (see equation (\ref{Kof})). The independence of the
contracted invariants is proved directly. It follows from equation
(\ref{ST}) that for any $2\leq j\leq N$ following identities hold:
\begin{equation}
\fl \frac{\partial ^{3}C_{j}}{\partial e_{1\mu }\partial e_{1\nu
}\partial e_{1\lambda }}=\frac{\partial ^{3}C_{j}}{\partial
e_{1\mu }\partial e_{1\nu }\partial f_{1\lambda }}=\frac{\partial
^{3}C_{j}}{\partial f_{1\mu
}\partial f_{1\nu }\partial e_{1\lambda }}=\frac{\partial ^{3}C_{j}}{%
\partial f_{1\mu }\partial f_{1\nu }\partial f_{1\lambda }}=0.
\end{equation}
That means that any invariant\footnote{Of degree $d\geq 2$.} of
$\frak{u}\left( p,q\right) $ can be written as a quadratic
polynomial in the variables $\left\{ e_{1\mu },f_{1\mu }\right\} $
with coefficients in the polynomial ring $\mathbb{R}\left[ e_{\mu
\nu },f_{\mu \nu }\right] _{2\leq \mu ,\nu \leq N}$. Moreover, it
is
straightforward to verify that no function depending only on the variables $%
\left\{ e_{\mu \nu },f_{\mu \nu }\right\} $ with $\mu ,\nu \neq 1$
is an invariant of $I\frak{u}\left( p,q\right) $, because of the
the action of the reductive part $\frak{u}\left( p-1,q\right) $
over the representation given in (\ref{Dar}). From these facts it
follows that the functions $D_{k}$ are algebraically independent
homogeneous polynomials, and therefore their symmetrization are
independent Casimir invariants \cite{AA}. By algebraic
manipulation, we can reduce and simplify the determinant
(\ref{Def}) in order to obtain an expression that does not involve
anymore the contraction parameter $t$. This is done applying the
usual determinantal methods, like the Laplace expansion. After
some decompositions and computations, we obtain that
\begin{equation}
\Delta =\left| B_{p,q}-\lambda \mathrm{Id}_{N}\right| +\lambda
\left| (B_{p,q})_{11}-\lambda \mathrm{Id}_{N-1}\right| ,
\label{Pol1}
\end{equation}
where
\begin{equation}
\fl B_{p,q}:=\left(
\begin{array}{ccccc}
0 & .. & g_{jj}\left( f_{1j}-ie_{1j}\right)  & .. &
g_{N+1,N+1}\left(
f_{1,N+1}-ie_{1,N+1}\right)  \\
: &  & : &  & : \\
\left( ie_{1j}+f_{1j}\right)  & .. & g_{jj}f_{jj} & .. &
g_{N+1,N+1}\left(
f_{j,N+1}-ie_{j,N+1}\right)  \\
: &  & : &  & : \\
(ie_{1,N+1}+f_{1,N+1}) & .. & g_{jj}\left(
ie_{j,N+1}+f_{j,N+1}\right)  & .. & g_{N+1,N+1}f_{N+1,N+1}
\end{array}
\right) ,  \label{Pol2}
\end{equation}
and $(B_{p,q})_{11}$ is the minor obtained deleting the first row
and column. Now, replacing $e_{1j}$ by $r_{j}$ and $f_{1j}$ by
$q_{j}$, equation (\ref{Pol2}) gives the Casimir operators of
$I\frak{u}\left( p-1,q\right) $ over the usual basis $\left\{
E_{\mu \nu },F_{\mu \nu },Q_{\rho },R_{\rho }\right\} $. The
advantage of formula (\ref{Pol2}) with respect to equation
(\ref{Def}) is that the invariants are expressed in the usual
basis of $I\frak{u}(p,q)$, without taking into account that this
algebra arises as contraction of the unitary algebra
$\frak{u}(p+1,q)$, and therefore avoiding the computation of
limits.

\section{A comment concerning the special pseudo-unitary algebras}

Formula (\ref{ST}) can be easily adapted to the special algebras.
For convenience we take the Cartan subalgebra spanned by the
vectors
$H_{\mu}=g_{\mu+1,\mu+1}F_{\mu\mu}-g_{\mu\mu}F_{\mu+1,\mu+1}$ for
$\mu=1..p+q-1$. The centre of $\frak{u}(p,q)$ is obviously
generated by $g^{\mu\mu}F_{\mu\mu}$, which coincides with the
first order invariant obtained previously.

\begin{proposition}
For $N\geq2$, a maximal set of independent Casimir invariants of
$\frak{su}\left(  p,q\right)  $ (where $p+q=N$) is given by the coefficients
$D_{k}$ of the characteristic polynomial $\left|  iA_{p,q}-\lambda
{\rm Id}_{N}\right|  =\lambda^{N}+\sum_{k=2}^{N}D_{k}\lambda^{N-k}$, where
\begin{equation}
\fl A_{p,q}=\left(
\begin{array}
[c]{ccccc}%
-iY_{1} & .. & -g_{\mu\mu}(e_{1\mu}+if_{1\mu}) & .. & -g_{NN}\left(
e_{1N}+if_{1N}\right)  \\
: &  & : &  & :\\
e_{1\mu}-if_{1\mu} & .. & -iY_{\mu} & .. & -g_{NN}\left(  e_{\mu N}+if_{\mu
N}\right)  \\
: &  & : &  & :\\
e_{1N}-if_{1N} & .. & g_{\mu\mu}\left(  e_{\mu N}-if_{\mu N}\right)   & .. &
-iY_{N}%
\end{array}
\right)  , \label{Tv}
\end{equation}
and
\begin{equation}
Y_{\mu}=\sum_{\nu=1}^{\mu-1}\frac{-\nu}{N}g_{\nu\nu}g_{\nu+1,\nu+1}h_{\nu}+\sum_{\nu=\mu}%
^{N-1}\frac{N-\nu}{N}g_{\nu\nu}g_{\nu+1,\nu+1}h_{\nu},\;1\leq\mu\leq N.
\end{equation}
Moreover, $\deg(D_{k})=k$ for $k=2..p+q=N$.
\end{proposition}

The interest of the preceding formula (\ref{Tv}) is its
application to obtain closed expressions for the Casimir operators
of some contractions of $\frak{su}(p,q)$. Consider for instance
the automorphism
\begin{equation}
\fl
\begin{array}
[c]{lll}%
\Phi\left( H_{\mu}\right)  =H_{\mu}\; (\mu\geq 1), & \Phi\left(
E_{1\sigma}\right)  =\frac{1}{t}E_{1\sigma}\; (\sigma\geq 2), \\
\Phi\left(F_{1\sigma}\right)  =\frac{1}{t}F_{1\sigma}\;
(\sigma\geq2), & \Phi\left(  E_{\mu\nu}\right)  =E_{\mu\nu}, &
\Phi\left(  F_{\mu\nu}\right) =F_{\mu\nu},\;\mu,\nu\neq1.
\end{array} \label{K10}
\end{equation}

For any pair of elements $X,Y$ of the algebra
$\lim_{t\rightarrow\infty}\Phi^{-1}\left[ \Phi(X),\Phi(Y)\right] $
exists, and defines a Lie algebra which will be denoted by
$K\frak{su}(p-1,q)$.\footnote{The Levi part, which is isomorphic
to $\frak{su}(p-1,q)$, is generated by $\lbrace
H_{\mu},E_{\mu\nu},F_{\mu\nu}\rbrace_{2\leq\mu<\nu}$.} It follows
at once from (\ref{K10}) that $K\frak{su}(p-1,q)$ is isomorphic to
an extension of degree one of the inhomogeneous algebra
$I\frak{su}(p-1,q)$. In fact, observe that the action of the
Cartan subalgebra $\frak{H}$ over the $E_{\mu,\nu}$ and
$F_{\mu,\nu}$ is not changed by the contraction. If we consider
the element $H^{\prime}\in\frak{H}$ defined by
\begin{equation}
H^{\prime}=\sum_{\mu=1}^{p+q-1}(p+q-\mu)\prod_{\rho\neq \mu,\mu+1}g_{\rho\rho}H_{\mu}, \label{vw2}
\end{equation}
it is not difficult to see that taking the new basis of $\frak{H}$
generated by $\left\{H^{\prime},H_{2},..,H_{p+q-1}\right\}$ we
have $\left[ H^{\prime },E_{\mu \nu }\right] =\left[ H^{\prime },F_{\mu \nu }%
\right] =0$ and
\begin{equation}
\left[ H^{\prime },E_{1\mu }\right] =-2(p+q)\prod_{\rho
=1}^{p+q}g_{\rho \rho }F_{1\mu },\quad \left[ H^{\prime },F_{1\mu
}\right] =2(p+q)\prod_{\rho =1}^{p+q}g_{\rho \rho }E_{1\mu }
\end{equation}
for $2\leq \mu <\nu $. This shows that
$\left[H^{\prime},\frak{su}(p-1,q)\right]=0$. Using this fact, it
is straightforward to verify that the linear mapping $\varphi:
I\frak{su}(p-1,q)\longrightarrow I\frak{su}(p-1,q)$ defined by
$\varphi(X)=\left[H^{\prime},X\right]$ is a derivation of the
inhomogeneous algebra $I\frak{su}(p-1,q)$. This result allows us
to determine the Casimir operators of the extension using the
preceding formuale. The proof is completely analogous to that of
proposition 2.

\begin{proposition}
For $N=p+q\geq2$, a maximal set of independent Casimir invariants
of $K\frak{su}\left(p-1,q\right)$ over the basis
$\left\{H^{\prime},H_{2},..,H_{N-1},E_{\mu\nu},F_{\mu\nu}\right\}$
 is given by the coefficients $W_{k}$ of the polynomial
\begin{equation}
\left|iC_{p,q}-\lambda {\rm Id}_{N}\right|+\lambda \left|i(C_{p,q})_{11}
-\lambda {\rm Id}_{N-1}\right|=\sum_{k=2}^{N}W_{k}\lambda^{N-k},\label{KK}
\end{equation}
where
\begin{equation}
\fl C_{p,q}=\left(
\begin{array}
[c]{ccccc}%
0 & .. & -g_{\mu\mu}(e_{1\mu}+if_{1\mu}) & .. & -g_{NN}\left(
e_{1N}+if_{1N}\right)  \\
: &  & : &  & :\\
e_{1\mu}-if_{1\mu} & .. & -iY_{\mu} & .. & -g_{NN}\left(  e_{\mu N}+if_{\mu
N}\right)  \\
: &  & : &  & :\\
e_{1N}-if_{1N} & .. & g_{\mu\mu}\left(  e_{\mu N}-if_{\mu N}\right)   & .. &
-iY_{N}%
\end{array}
\right)  , \label{Tv1}
\end{equation}
and
\begin{equation}
\fl
Y_{\mu}=-\prod_{\rho=1}^{N}g_{\rho\rho}\frac{h^{\prime}}{N(N-1)}+\sum_{\nu=2}^{\mu-1}
\frac{1-\nu}{N-1}g_{\nu\nu}g_{\nu+1,\nu+1}h_{\nu}+\sum_{\nu=\mu}^{N-1}\frac{N-\nu}{N-1}g_{\nu\nu}g_{\nu+1,\nu+1}h_{\nu}
\end{equation}
for $2\leq \mu\leq N$. Moreover, $\deg(W_{k})=k$ for $k=2..p+q=N$.
\end{proposition}

The functions $W_{k}$ have a structural property that will be of
interest in the analysis of the Casimir operators of inhomogeneous algebra $I\frak{su}%
\left( p,q\right) $.

\begin{proposition}
For any $N\geq 2$, the functions $W_{k}$ of (\ref{KK}) satisfy the
 equation
\begin{equation}
N\left( N-1\right) \frac{\partial W_{k}}{\partial h^{\prime }}-\left(
N+1-k\right) \prod_{\rho =1}^{N}g_{\rho \rho }W_{k-1}=0. \label{Fun}
\end{equation}
\end{proposition}

\begin{proof}
By (\ref{KK}), the $W_{k}$ are the result of a combination of the
characteristic polynomial of the matrix $iC_{p,q}$ and the minor
$i\left( C_{p,q}\right) $ obtained deleting the first row and
column. Therefore, using the expansion properties of determinants,
it follows easily that any $W_{k}$ is obtained from the sum of the
minors of order $k$ of $iC_{p,q}$ that contain the first column
and row, i.e.,
\begin{equation}
W_{k}=\sum_{2<\mu _{1}..<\mu _{k-1}}M\left( 1,\mu _{1},..,\mu _{k-1}\right) ,
\end{equation}
where $M\left( 1,\mu _{1},..,\mu _{k-1}\right)$ is the matrix given by:
\begin{equation}
\fl \left(
\begin{array}{cccc}
0 & -g_{\mu _{1}\mu _{1}}(ie_{1\mu _{1}}-f_{1\mu _{1}}) & .. & -g_{\mu
_{k-1},\mu _{k-1}}\left(i e_{1\mu _{k-1}}-f_{1\mu _{k-1}}\right)  \\
(ie_{1\mu _{1}}+f_{1\mu _{1}}) & Y_{\mu _{1}} & .. & -g_{\mu _{k-1},\mu
_{k-1}}\left( ie_{\mu _{1}\mu _{k-1}}-f_{\mu _{1}\mu _{k-1}}\right)  \\
: & : &  & : \\
\left(i e_{1\mu _{k-1}}+f_{1\mu _{k-1}}\right)  & g_{\mu _{1}\mu _{1}}\left(
ie_{\mu _{1}\mu _{k-1}}+f_{\mu _{1}\mu _{k-1}}\right)  & .. & Y_{\mu
_{k-1}}
\end{array}
\right) .
\end{equation}
Now the variable $h^{\prime }$ is contained only in the elements
$Y_{\mu }$ of the diagonal, which implies that
\begin{eqnarray}
\fl \frac{\partial W_{k}}{\partial h^{\prime }} =-\sum_{2<\mu _{1}..<\mu _{k-1}}%
\frac{\partial M\left( 1,\mu _{1},..,\mu _{k-1}\right) }{\partial h^{\prime }%
}=-\sum_{2<\mu _{1}..<\mu _{k-1}}\sum_{j=\mu _{1}}^{\mu _{k-1}}\frac{\partial
Y_{j}}{\partial h^{\prime }}M\left( 1,\mu _{1},..,\mu _{k-2}\right) = \nonumber \\
\fl =\frac{\left( N+1-k\right) }{N\left( N-1\right) }\prod_{\rho
=1}^{N}g_{\rho \rho }\sum_{2<\mu _{1}..<\mu _{k-2}}M\left( 1,\mu _{1},..,\mu
_{k-2}\right) =\frac{\left( N+1-k\right) }{N\left( N-1\right) }\prod_{\rho
=1}^{N}g_{\rho \rho }W_{k-1}.
\end{eqnarray}
\end{proof}

Since $K\frak{su}(p-1,q)$ is a contraction of $\frak{su}(p,q)$, by
 transitivity of contractions it follows that $K\frak{su}(p-1,q)$
also contracts onto $I\frak{su}(p-1,q)\oplus <H^{\prime}>$, and we
obtain the chain of contractions
\begin{equation}
\frak{su}(p,q) \rightsquigarrow K\frak{su}(p-1,q) \rightsquigarrow I\frak{su}(p-1,q)\oplus <H^{\prime}>.
\end{equation}

The question is whether the procedure developed to compute the
Casimir invariants can be adapted to the special inhomogeneous
case. The following example shows that in general, the contraction
of the invariants forming a fundamental set of invariants of
$\frak{su}(p,q)$ (or $K\frak{su}(p-1,q)$) does not generate a
complete set of invariants of the contraction. Consider the Lie
algebra $\frak{su}\left(3,1\right)  $. This algebra has three
invariants $C_{2},C_{3},C_{4}$ which are obtained from the
characteristic polynomial of the matrix
\begin{equation}
\fl D=\left(
\begin{array}
[c]{cccc}%
\frac{3}{4}h_{1}+\frac{1}{2}h_{2}-\frac{1}{4}h_{3} & -ie_{12}+f_{12} &
-ie_{13}+f_{13} & ie_{14}-f_{14}\\
ie_{12}+f_{12} & -\frac{1}{4}h_{1}+\frac{1}{2}h_{2}-\frac{1}{4}h_{3} &
-ie_{23}+f_{23} & ie_{24}-f_{24}\\
ie_{13}+f_{13} & ie_{23}+f_{23} & -\frac{1}{4}h_{1}-\frac{1}{2}h_{2}-\frac
{1}{4}h_{3} & ie_{34}-f_{34}\\
ie_{14}+f_{14} & ie_{24}+f_{24} & ie_{34}+f_{34} & -\frac{1}{4}h_{1}-\frac
{1}{2}h_{2}+\frac{3}{4}h_{3}%
\end{array}
\right).\label{su31}
\end{equation}
We obtain the inhomogeneous algebra $I\frak{su}(2,1)$ from the
contraction determined by the automorphism
\begin{equation}
\fl
\begin{array}
[c]{lll}%
\Phi\left( H_{1}\right)  =\frac{1}{t}H_{1}, & \Phi\left(  F_{\mu
}\right)  =F_{\mu}\;(\mu\geq2), & \Phi\left(  E_{1\sigma}\right)
=\frac{1}{t}E_{1\sigma}\; (\sigma\geq 2),\\
\Phi\left(F_{1\sigma}\right)
=\frac{1}{t}F_{1\sigma}\;(\sigma\geq2), & \Phi\left(
E_{\mu\nu}\right)  =E_{\mu\nu}, & \Phi\left(  F_{\mu\nu}\right)
=F_{\mu\nu},\;\mu,\nu\neq1.
\end{array} \label{K11}
\end{equation}
However, since we are rescaling also a generator of the Cartan
subalgebra, it will follow that the contraction of the  Casimir
operators $C_{i}(h_{\mu},e_{\mu\nu},f_{\mu\nu})$ for $i=2,3,4$ are
functions of the quadratic operator
$I_{2}=g_{\mu\mu}(e_{1\mu}^2+f_{1\mu}^2)$ of $I\frak{su}(2,1)$ and
$h_{1}$\footnote{Observe that in the contraction this becomes a
central element.}. More specifically, for $k=2,3,4$ we have:
\begin{equation}
\lim_{t\rightarrow\infty}\frac{1}{t^{k}}C_{k}=-\alpha_{k}I_{2}h_{1}^{k-2}
,\label{kt1}
\end{equation}
where $\alpha_{2}=1, \alpha_{3}=\frac{1}{2},
\alpha_{4}=\frac{3}{16}$. Therefore the characteristic polynomial
of the matrix (\ref{su31}) does not provide a maximal set of
invariants of $I\frak{su}(2,1)$. However, it is easy to obtain an
independent invariant using (\ref{kt1}). This third invariant,
which must depend on all variables of $I\frak{su}(2,1)$
\cite{Ca3}, can be obtained for example contracting the invariant
$C_{2}C_{4}-\frac{1}{4}C_{3}^{2}$, and provides a sixth order
Casimir operator of $I\frak{su}(2,1)$ which is not a power of
$I_{2}$. A similar result holds if we start from the invariants of
$K\frak{su}\left(2,1\right)$. We can however proceed differently.
Instead of contracting the invariants $W_{k}$,  we evaluate them
in the system (\ref{sys}) corresponding to $I\frak{su}(2,1)$ over
the basis $\left\{H_{2},H_{3},E_{\mu \nu },F_{\mu \nu }\right\} $.
We obtain

\begin{equation}
\fl
\begin{tabular}{lll}
$\widehat{H}_{\mu }\left( W_{i}\right) =0\; (\mu=2,3),$ &
$\widehat{E}_{\mu \nu }\left( W_{2}\right) =0$, &
$\widehat{F}_{\mu
\nu}\left( W_{2}\right) =0$, \\
$\widehat{E}_{\mu \nu }\left( W_{3}\right) =-\frac{4}{3}f_{\mu \nu
}W_{2},$ & $\widehat{F}_{\mu \nu }\left( W_{3}\right)
=\frac{4}{3}e_{\mu \nu }W_{2},$
& $\widehat{E}_{\mu \nu }\left( W_{4}\right) =-\frac{2}{3}f_{\mu \nu }W_{3}$.%
\end{tabular}
\end{equation}
Now let $u=W_{3}$ and $v=W_{4}$. For any $X\in \left\{ E_{\mu \nu
},F_{\mu \nu }\right\} $ we have $\frac{\widehat{X}\left( v\right) }{\widehat{X}\left( u\right) }=\frac{W_{3}}{%
2W_{2}}=\frac{u}{2W_{2}}$. If we now consider the differential
equation
\begin{equation}
\frac{\partial F}{\partial u}+\frac{u}{2W_{2}}\frac{\partial F}{\partial v}=0, \label{DGL}
\end{equation}
it follows easily that the solutions are generated by the function
$(W_{2}v-\frac{1}{4}u^{2})W_{2}^{-1}=(W_{2}W_{4}-\frac{1}{4}W_{3}^{2})W_{2}^{-1}$.
Since $W_{2}$ is already an invariant of $I\frak{su}\left(
2,1\right) $, this implies that $I=4W_{2}W_{4}-W_{3}^{2}$ is an
invariant of the algebra of degree six. A short computation shows
that it coincides with
$\lim_{t\rightarrow\infty}\frac{1}{t^{4}}(C_{2}C_{4}-\frac{1}{4}C_{3}^{2})$.
This fact is not casual and can be formulated in any dimension.

\begin{theorem}
Let $p+q=N\geq 2$ and $W_{k}$ be the invariants of the Lie algebra
$K\frak{su}(p-1,q)$. A function $G(W_{2},..,W_{N})$ is an
invariant of the inhomogeneous Lie algebra $I\frak{su}(p-1,q)$ if
and only if
\begin{equation*}
\frac{\partial{G(W_{2},..,W_{N})}}{\partial h^{\prime}}=0.
\end{equation*}
\end{theorem}

\begin{proof}
Let us consider the
bases of $K\frak{su}(p-1,q)$ and $I\frak{su}(p-1,q)$ used previously,and
let $\left\{ \widehat{H^{\prime}},\widehat{H}_{\mu },\widehat{E}_{\mu \nu },
\widehat{F}_{\mu \nu }\right\} $ and $\left\{ \widetilde{H^{\prime }},%
\widetilde{H}_{\mu },\widetilde{E}_{\mu \nu },\widetilde{F}_{\mu \nu
}\right\} $ be the representation by differential operators of $K\frak{su}%
\left( p-1,q\right) $, respectively $I\frak{su}\left(p-1,q\right)
\oplus <H^{\prime }>$. Since $K\frak{su}\left(p-1,q\right) $ is an
extension by a derivation of $I\frak{su}\left( p-1,q\right) $
acting trivially on the Levi part, for the invariants $W_{k}$ we
obtain the relations
\begin{eqnarray}
\widehat{H}_{\mu }\left( W_{\lambda }\right)  &=&\widetilde{H}_{\mu }\left(
W_{\lambda }\right) =0,\;1\leq \mu \leq N-1 \\
\widehat{E}_{\mu \nu }\left( W_{\lambda }\right)  &=&\widetilde{E}_{\mu \nu
}\left( W_{\lambda }\right) =0,\;2\leq \mu ,v\leq N \\
\widehat{F}_{\mu \nu }\left( W_{\lambda }\right)  &=&\widetilde{F}_{\mu \nu
}\left( W_{\lambda }\right) =0,\;2\leq \mu ,v\leq N \\
\widehat{E}_{1\mu }\left( W_{\lambda }\right)  &=&\widetilde{E}_{1\mu
}\left( W_{\lambda }\right) +2N\prod_{\rho =1}^{N}g_{\rho \rho }\frac{%
\partial W_{\lambda }}{\partial h^{\prime }}f_{1\mu}=0,\;2\leq \mu \leq N \\
\widehat{F}_{1\mu }\left( W_{\lambda }\right)  &=&\widetilde{F}_{1\mu
}\left( W_{\lambda }\right) -2N\prod_{\rho =1}^{N}g_{\rho \rho }\frac{%
\partial W_{\lambda }}{\partial h^{\prime }}e_{1\mu}=0,\;2\leq \mu \leq N.
\end{eqnarray}
For $W_{2}$ we obviously have $\frac{\partial W_{2}}{\partial h^{\prime }}%
=0$, we thus recover the quadratic invariant. For any $\lambda \geq 3$ the
previous system, combined with proposition 5, provides the following expressions for any $%
2\leq \mu \leq N$:
\begin{eqnarray}
\widetilde{E}_{1\mu }\left( W_{\lambda }\right)  &=&-2N\prod_{\rho
=1}^{N}g_{\rho \rho }\frac{\partial W_{\lambda }}{\partial h^{\prime }}=-
\frac{2\left( N+1-k\right) }{\left( N-1\right) }W_{\lambda -1}f_{1\mu}, \label{D1}\\
\widetilde{F}_{1\mu }\left( W_{\lambda }\right)  &=&2N\prod_{\rho
=1}^{N}g_{\rho \rho }\frac{\partial W_{\lambda }}{\partial h^{\prime }}=
\frac{2\left( N+1-k\right) }{\left( N-1\right) }W_{\lambda -1}e_{1\mu}.\label{D2}
\end{eqnarray}

As a consequence of these identities, an invariant $W$ of $K\frak{su}\left( p-1,q\right) $ is also
an invariant of the inhomogeneous algebra $I\frak{su}\left( p-1,q\right) $
if and only if $\frac{\partial W}{\partial h^{\prime }}=0$.
\end{proof}

Observe that equations (\ref{D1}) and (\ref{D2}) can be used to
compute additional invariants of $I\frak{su}(p-1,q)$ using
differential equations of the type (\ref{DGL}). The fact that the
resulting function must be independent of $h^{\prime}$ will impose
some restrictions on the degree.

\section{On the applicability of the method}

Although the method to compute Casimir invariants by means of
characteristic polynomials has been developed for the Lie algebras
$\frak{u}(p,q)$ and their contractions, the argument is not
exclusive of the unitary case, and does not depend on the special
shape of their invariants. In this section we give matrix formulae
for the invariants of other inhomogeneous Lie algebras of physical
importance. More specifically, we apply the method to the
inhomogeneous pseudo-orthogonal Lie algebras, a double
inhomogeneous algebra and a semidirect product of a Heisenberg and
an orthogonal Lie algebra, which can both be obtained from
contractions of $I\frak{so}(p,q)$ and were studied in \cite{Hp}.

\subsection{Inhomogeneous pseudo-orthogonal Lie algebras}

The inhomogeneous pseudo-orthogonal algebra $I\frak{so}(p,q)$ with
$N=p+q$ is given by the  $\frac{1}{2}N(N+1)$ operators
$E_{\mu\nu}=-E_{\nu\mu},P_{\mu}$,  satisfying:
\begin{eqnarray*}
\left[ E_{\mu \nu },E_{\lambda \sigma }\right]  &=&g_{\mu \lambda
}E_{\nu \sigma }+g_{\mu \sigma }E_{\lambda \nu }-g_{\nu \lambda
}E_{\mu \sigma
}-g_{\nu \sigma }E_{\lambda \mu } \\
\left[ E_{\mu \nu },P_{\rho }\right]  &=&g_{\mu \rho }P_{\nu
}-g_{\nu \rho }P_{\mu },
\end{eqnarray*}
where $g={\rm diag}\left( 1,..,1,-1,..,-1\right)$. Moreover, in
\cite{De2} it was shown that
$\mathcal{N}(I\frak{so}(p,q))=\left[\frac{p+q+1}{2}\right]$.

\begin{proposition}
A maximal set of Casimir operators of $I\frak{so}(p,q)$ is given
by the coefficients $C_{k}$ of $P(T)$ defined by
\begin{equation}
P(T):=\left| B_{p,q}-T\mathrm{Id}_{N+1}\right| +T\left| (B_{p,q})_{(N+1,N+1)}-T\mathrm{Id%
}_{N}\right| ,  \label{Pol1}
\end{equation}
where
\begin{equation}
B_{p,q}:=\left(
\begin{array}{cccccc}
0 & .. & -g_{jj}e_{1j} & .. & -g_{NN}e_{1,N} & p_{1}T \\
: &  & : &  & : &  \\
e_{1j} & .. & 0 & .. & -g_{NN}e_{j,N} & p_{j}T \\
: &  & : &  & : &  \\
e_{1,N} & .. & g_{jj}e_{j,N} & .. & 0 & p_{N}T \\
-p_{1} &  & -g_{jj}p_{j} &  & -g_{NN}p_{N} & 0
\end{array}
\right)
\end{equation}
\end{proposition}

The proof is quite similar to that of proposition 2. Here we use
the contraction of $\frak{so}(p+1,q)$ over $I\frak{so}(p,q)$, as
done in \cite{Ro}. In contrast to the unitary case, where the
matrices involved are complex, here no problem of dependence in
the contracted invariants is observed.

\subsection{The double inhomogeneous algebra $AG(p,q)$}

The Lie algebra $AG\left(p,q\right) $ with $N=p+q+1$ is given by the $%
\frac{1}{2}\left( N+1\right)N$ operators $E_{\mu \nu }=-E_{\nu \mu
},P_{\mu },G_{\nu },P_{0}$, satisfying the nontrivial commutation
relations:
\begin{equation}
\fl \begin{tabular}{ll} $\left[ E_{\mu \nu },E_{\lambda \sigma
}\right] =g_{\mu \lambda }E_{\nu \sigma }+g_{\mu \sigma
}E_{\lambda \nu }-g_{\nu \lambda }E_{\mu \sigma }-g_{\nu \sigma
}E_{\lambda \mu },$ & $\left[ P_{0},Q_{\rho }\right]
=-g_{N-1,N-1}P_{\rho },$ \\
$\left[ E_{\mu \nu },P_{\rho }\right] =g_{\mu \rho }P_{\nu
}-g_{\nu \rho }P_{\mu },\;$ & $\left[ E_{\mu \nu },Q_{\rho
}\right] =g_{\mu \rho }Q_{\nu
}-g_{\nu \rho }Q_{\mu },$%
\end{tabular}
\end{equation}
where $1\leq \mu ,\nu ,\rho ,\leq N-1$ and $\left( g_{\mu \nu
}\right)$ is the preceding diagonal matrix. It can be easily shown
 that this algebra is obtained by an In\"{o}n\"{u}-Wigner
contraction of the inhomogeneous Lie algebra $I\frak{so}\left(
p+1,q\right) $, considering the automorphism
\begin{equation}
\begin{tabular}{ll}
$\Psi \left( E_{\mu \nu }\right) =E_{\mu \nu }\;\left( 1\leq \mu
,\nu \leq N-1\right) ,$ & $\Psi \left( E_{\mu N}\right)
=\frac{1}{t}E_{\mu
N}:=G_{\mu },$ \\
$\Psi \left( P_{\mu }\right) =\frac{1}{t}P_{\mu }\;\left( 1\leq
\mu \leq
N-1\right) ,$ & $\Psi \left( P_{N}\right) =P_{N}:=P_{0}.$%
\end{tabular}
\end{equation}
This contraction preserves the number of invariants, i.e.,
$\mathcal{N}(AG(p,q)=\left[\frac{p+q+1}{2}\right]$. Moreover, it
should be remarked that $AG(p,q)$ is a double inhomogeneous Lie
algebra with semidirect product structure
\begin{equation}
AG(p,q)=II\frak{so}(p,q)=\left(\frak{so}(p,q)\overrightarrow{\oplus}I_{N-1}\right)\overrightarrow{\oplus}I_{N},
\end{equation}
$I_{N}$ being the standard vector representation. In particular,
for $q=0$ we obtain the proper Galilei algebra $II\frak{so}(N-1)$.

\begin{proposition}
A maximal set of Casimir operators of $AG(p,q)$ is given by the
coefficients $C_{k}$ of the determinant $P(T)$ defined by
\begin{eqnarray}
\fl P\left( T\right) =\left|
\begin{array}{ccccccc}
-T & .. & -g_{\mu\mu}e_{1\mu} & .. & -g_{N-1,N-1}e_{1,N-1} &
-g_{N,N}q_{1} & p_{1}T
\\
: &  & : &  & : &  &  \\
e_{1\mu} & .. & -T & .. & -g_{N-1,N-1}e_{\mu,N-1} & -g_{N,N}q_{\mu} & p_{\mu}T \\
: &  & : &  & : &  &  \\
e_{1,N-1} & .. & g_{\mu\mu}e_{\mu,N-1} & .. & -T & -g_{N,N}q_{N-1} & p_{N-1}T \\
q_{1} &  & g_{\mu\mu}q_{\mu} &  & g_{N-1,N-1}q_{N-1} & 0 & 0 \\
-p_{1} &  & -g_{\mu\mu}p_{\mu} &  & -g_{N-1,N-1}p_{N-1} & 0 & 0
\end{array}
\right| + \nonumber \\
\left|
\begin{array}{cccccc}
-T & .. & 0 &..  & 0& p_{1}T \\
:&  &: &  & : & : \\
0& .. & -T& .. & 0 & p_{\mu}T \\
:&  & : & & : & : \\
0& .. & 0 & ..& -T & p_{N-1}T \\
-p_{1} & .. & -g_{\mu\mu}p_{\mu} & .. & -g_{N-1,N-1}p_{N-1} & 0
\end{array}
\right|.
\end{eqnarray}
\end{proposition}

\subsection{The Lie algebra $C(p,q)$}

The Lie algebra $C\left(p,q\right) $ with $N=p+q+1$ is also
obtained from a contraction of $I\frak{so}(p+1,q)$ by means of the
automorphism defined by
\begin{equation}
\begin{tabular}{ll}
$\Psi \left( E_{\mu \nu }\right) =E_{\mu \nu }\;\left( 1\leq \mu
,\nu \leq N-1\right) ,$ & $\Psi \left( E_{\mu N}\right)
=\frac{1}{t}E_{\mu
N}:=G_{\mu },$ \\
$\Psi \left( P_{\mu }\right) =\frac{1}{t}P_{\mu }\;\left( 1\leq
\mu \leq
N-1\right) ,$ & $\Psi \left( P_{N}\right) =\frac{1}{t^{2}}P_{N}:=P_{0}.$%
\end{tabular}
\end{equation}
Over the basis $\left\{E_{\mu \nu },P_{\mu },G_{\nu
},P_{0}\right\}$, the nontrivial commutation relations are:
\begin{equation}
\fl \begin{tabular}{ll} $\left[ E_{\mu \nu },E_{\lambda \sigma
}\right] =g_{\mu \lambda }E_{\nu \sigma }+g_{\mu \sigma
}E_{\lambda \nu }-g_{\nu \lambda }E_{\mu \sigma }-g_{\nu \sigma
}E_{\lambda \mu },$ & $\left[ P_{\rho},Q_{\rho }\right]
=-g_{\rho\rho}P_{0},$ \\
$\left[ E_{\mu \nu },P_{\rho }\right] =g_{\mu \rho }P_{\nu
}-g_{\nu \rho }P_{\mu },\;$ & $\left[ E_{\mu \nu },Q_{\rho
}\right] =g_{\mu \rho }Q_{\nu
}-g_{\nu \rho }Q_{\mu },$%
\end{tabular}
\end{equation}
where $1\leq \mu ,\nu ,\rho ,\leq N-1$. This algebra is isomorphic
to the semidirect product of $\frak{so}(p,q)$ and the Heisenberg
Lie algebra $\frak{h}_{N}$, and therefore can be written as
\begin{equation}
C(p,q)=II^{\prime}\frak{so}(p,q)=\left(\frak{so}(p,q)\overrightarrow{\oplus}I_{N-1}^{\prime}\right)\overrightarrow{\oplus}I_{N},
\label{car}
\end{equation}
where the orthogonal algebra acts on $I_{N}$ through the vector
representation and on $I_{N-1}^{\prime}$ through the
contragredient of the vector representation. For $q=0$ we obtain
the generalization of the proper Carroll algebra
$C(p,0)=II^{\prime}\frak{so}(N-1)$. In particular it follows from
(\ref{car}) that
$\mathcal{N}(C(p,q))=\left[\frac{p+q+1}{2}\right]$.

\begin{proposition}
A maximal set of Casimir operators of $C(p,q)$ is given by the
coefficients $C_{k}$ of the determinant $P(T)$ defined by
\begin{equation}
\fl P\left( T\right) =\left|
\begin{array}{ccccccc}
-T & .. & -g_{\mu\mu}e_{1\mu} & .. & -g_{N-1,N-1}e_{1,N-1} &
-g_{N,N}q_{1} & p_{1}T
\\
: &  & : &  & : &  &  \\
e_{1\mu} & .. & -T & .. & -g_{N-1,N-1}e_{\mu,N-1} & -g_{N,N}q_{\mu} & p_{\mu}T \\
: &  & : &  & : &  &  \\
e_{1,N-1} & .. & g_{\mu\mu}e_{\mu,N-1} & .. & -T & -g_{N,N}q_{N-1} & p_{N-1}T \\
q_{1} &  & g_{\mu\mu}q_{\mu} &  & g_{N-1,N-1}q_{N-1} & 0 & p_{0}T \\
-p_{1} &  & -g_{\mu\mu}p_{\mu} &  & -g_{N-1,N-1}p_{N-1} &
-g_{N,N}p_{0} & 0
\end{array}
\right|.
\end{equation}
\end{proposition}

\section{Concluding remarks}

We have seen that, in analogy with the classical algebras, the
Casimir operators of inhomogeneous pseudo-unitary Lie algebras
$I\frak{u}(p,q)$ can be computed using only characteristic
polynomials, avoiding the use of the enveloping algebra
\cite{De2}. This formula is derived from the In\"on\"u-Wigner
contraction
\begin{equation}
\frak{u}(p,q) \rightsquigarrow I\frak{u}(p-1,q)\oplus\mathbb{R},
\end{equation}
and allows to give a closed matrix expression for the invariants
of the inhomogeneous algebras. The method can be enlarged to cover
other contractions which are not necessarily inhomogeneous
algebras.  After adapting the formula to the special
pseudo-unitary algebras $\frak{su}(p,q)$, we analyze the problem
of computing the invariants for the inhomogeneous special algebras
$I\frak{su}(p-1,q)$ applying the same procedure. However, the
contraction of the invariants of $\frak{su}(p,q)$ points out that
the contraction of independent Casimir invariants does not
necessarily lead to independent Casimir operators of the
contraction. This fact is the consequence of altering the Cartan
subalgebra in the contraction. To solve this problem, special
pseudo-unitary algebras are contracted onto an extension
$K\frak{su}(p,q)$ of the inhomogeneous algebra $I\frak{su}(p,q)$.
The invariants $W_{k}$ of this extension are shown to satisfy a
functional equation (see (\ref{Fun})) that can be used to compute
the invariants of $I\frak{su}(p,q)$ by purely analytical methods.
This procedure enlarges the extension method introduced in
\cite{Ca33} for the special affine algebras
$\frak{sa}(n,\mathbb{R})$ and other algebras having only one
Casimir operator.

Further, we have exhibited examples that show the validity of the
argument when applied to other Lie algebras different from unitary
algebras. In particular, matrix formulae for the Casimir operators
of the inhomogeneous pseudo-orthogonal Lie algebras and two
In\"on\"u-Wigner contractions that naturally generalize the
generalized Galilei and Carroll algebras have been given. The
latter contractions of $I\frak{so}(p+1,q)$ preserve the number of
invariants \cite{Hp}. Moreover, the structure of $C(p,q)$ shows
that the procedure can be still valid for contractions of Lie
algebras that are no more inhomogeneous.

\bigskip

\section*{Acknowledgements}
The author expresses his gratitude to the referees for drawing the
attention to some mistakes, as well as for valuable suggestions
that helped to improve the manuscript.\newline  During the
preparation of this work, the author was partially supported by
the research project PR1/05-13283 of the U.C.M.

\end{document}